\documentstyle[11pt,aasms4,epsf]{article}

\def\ea{{\it et al.}~}
\def\etal{{\it et al.~}}
\def\eg{{\it e.g.,~}}
\def\ie{{\it i.e.,~}}

\def \msol {\rm{M}$_\odot$}
\def \mdot {\rm{M}$_\odot$~yr$^{-1}$}
\def \kms{km~$\rm{s}^{-1}$}
\def \cc{$\rm{cm}^{-3}$}

\newcommand{\be}{\begin{equation}}
\newcommand{\ee}{\end{equation}}
\newcommand{\vpeo}{\mbox{${v_{\perp o}}$}}
\newcommand{\vpei}{\mbox{${v_{\perp i}}$}}
\newcommand{\vpao}{\mbox{${v_{\parallel o}}$}}
\newcommand{\vpai}{\mbox{${v_{\parallel i}}$}}
\newcommand{\upe}{\mbox{${u_{\perp}}$}}
\newcommand{\upa}{\mbox{${u_{\parallel}}$}}
\newcommand{\vpe}{\mbox{${v_{\perp}}$}}
\newcommand{\vpa}{\mbox{${v_{\parallel}}$}}

\begin{document}

\title{Supernova 1987A: Rotation and a Binary Companion}

\author{T.J.B. Collins}
\affil{Laboratory for Laser Energetics,\\
University of Rochester, 250 E. River Rd., Rochester, NY 14623-1299}

\author{Adam Frank}
\affil{Dept. of Physics and Astronomy and C. E. K. Mees Observatory,\\
University of Rochester, Rochester, NY 14627-0171}

\author{J.E. Bjorkman}
\affil{Dept. of Physics and Astronomy, Ritter Observatory, \\
University of Toledo, Toledo, OH 43606-3390}

\author{Mario Livio} \affil{Space Science Telescope Institute, 3700 San
Martin Dr. Baltimore, MD 21218}

\begin{center}{\tt Draft \today}\end{center}

\begin{abstract}
In this paper we provide a possible link between the structure of the
bipolar nebula surrounding SN1987A and the properties of its
progenitor star. A Wind Blown Bubble (WBB) scenario is employed, in
which a fast, tenuous wind from a Blue Supergiant expands into a slow,
dense wind, expelled during an earlier Red Supergiant phase. The
bipolar shape develops due to a pole-to-equator density contrast in
the slow wind (\ie the slow wind forms a {\it slow torus}).  We use
the Wind Compressed Disk (WCD) model of Bjorkman \& Cassinelli (1992)
to determine the shape of the slow torus. In the WCD scenario, the
shape of the torus is determined by the rotation of the progenitor
star.  We then use a self-similar semi-analytical method for wind
blown bubble evolution to determine the shape of the resulting bipolar
nebula.

We find that the union of wind-compressed-disk and
bipolar-wind-blown-bubble models allows us to recover the salient
properties of SN1987A's circumstellar nebula.  In particular, the
size, speed and density of SN1987A's inner ring are easily reproduced
in our calculations.  An exploration of parameter space shows that the
red supergiant progenitor must have been rotating at $\gtrsim0.3$ of
its breakup speed.  We conclude that the progenitor was most likely
spun up by a merger with a binary companion. Using a simple model for
the binary merger we find that the companion is likely to have had a
mass $\gtrsim 0.5M_\odot$.

\end{abstract}

\section{Introduction}

The rings surrounding SN1987A (Burrows \ea 1995) are one example of the
ubiquitous phenomenon known as a bipolar outflow.  These hypersonic,
figure-{\bf 8} shaped nebulae occur in almost all forms of evolved
stars, regardless of their mass (\eg Planetary Nebulae: Manchado \ea
1996; Schwartz, Corradi \& Melnick 1992; Balick 1987,  Luminous Blue
Variables: Nota \ea 1995; and see Livio 1997 for a review).

Considerable progress has been made in understanding bipolar outflows
resulting from interacting stellar winds.  Analytical and numerical
studies have recovered many of the observed features of these outflows
through the {\em Generalized Wind Blown Bubble} (GWBB) paradigm (Kwok
\& West 1984; see Frank 1998 for a review). In this scenario, a fast
wind from a central source expands into a highly aspherical ({\em
i.e.}, toroidal) environment. The interaction of the wind and its
environment produces an expanding bubble bounded by strong shocks. The
bubble's velocity is highest in the direction of lowest density. As a
result, the density gradient in the environment (the {\em slow torus}),
establishes a preferred axis for the bipolar lobes.  The GWBB paradigm
has been successfully applied to all forms of evolved-star bipolar
outflows. Models which include the relevant hydrodynamics and
microphysics have recovered the global morphology, kinematics and
ionization patterns in many planetary nebulae (PNe; Frank \& Mellema
1994, Mellema 1996, Dwarkadas, Chevalier \& Blondin 1995). Clear
correspondences also exist between GWBB models and the shapes and
kinematics of Wolf-Rayet (WR) nebulae (Garcia-Segura \& MacLow 1993),
LBVs like $\eta$ Carina (Frank, Balick \& Davidson 1995) and symbiotic
stars like R Aquarii (Henney \& Dyson 1992).

The success of the GWBB paradigm opens the possibility of using the
properties of a nebula to infer the history of the central star in
terms of its mass loss. We note that as of yet no studies have
attempted to make an explicit and quantitaive link between the
properties of the nebula and history of the star.  The intense scrutiny
applied to SN87A makes it a unique laboratory for studying the
connection between bipolar outflows and stellar evolution.  Such is the
goal of this paper.  The recent discovery of a bipolar outflow
surrounding Sher 25 (Brander \ea 1997), a star similar to the
progenitor of SN87A, also raises the possibility that SN87A is not an
isolated object, but defines a new class of bipolar outflows.

When the central ring of SN87A was discovered, it was quickly
interpreted as the waist of a bipolar wind-blown bubble (Luo \& McCray
1991, Wang \& Mazzuli 1992; a structure anticipated by Soker \& Livio
1989).  In these models the ambient medium was taken to be a toroidal
wind deposited by SN87A's progenitor in its Red Supergiant (RSG)
stage.  The bubble was subsequently inflated by a fast wind from the
star's penultimate incarnation as a Blue Supergiant (BSG). The first
numerical simulations of this process were carried out by Blondin \&
Lundqvist (1993: BL93) who used an {\em ad hoc} function to determine
the asphericity of the slow RSG wind.  BL93 were quite successful in
demonstrating the feasibility of the GWBB paradigm for SN87A. However,
the observed low expansion speed of the central ring ($v_\varpi
\approx 8$ \kms, Meaburn \ea 1995) presented a problem.  To create a
model with the correct kinematics, BL93 were forced to adopt very low
values of both the RSG and BSG wind velocities. In particular, their
value of the RSG wind speed, $v_{a} \sim 5$ \kms, ~seemed particularly
anomalous.  Canonical values for Red Giant winds are comparable to 20
km/s (Habing 1996).  BL93 were also forced to take an equator-to-pole
density contrast of $p = \rho_e/\rho_p =20$, which, at the time,
seemed large.  The size of $p$ was one reason cited by McCray \& Lin
1992 in their arguments that the ambient density distribution
represented a remnant protostellar disk rather than a stellar wind
(although more recently the protostellar disk idea has been abandoned,
McCray, private communication).  Martin and Arnett (1994, MA95) carried
out GWBB simulations similar to Blondin \& Lundqvist.  Using a
different {\em ad hoc} function to control the density asymmetry of
the RSG wind, MA95 confirmed Blondin \& Lundqvist's results, including
the need for low RSG wind speeds, $v_{a} < 10$ km/s. We note that both
BL93 and MA95 assumed that the the RSG wind speed was isotropic.

The discovery of additional upper and lower rings of SN87A both
confirmed and confused the image of SN87A's nebula as a bipolar
outflow.  While a number of different models for the the formation of
these outer rings have been proposed (\eg Burrows \ea, 1995; MA95;
Podsiadlowski, Fabian \& Stevens, 1991; Lloyd, O'Brian \& Kahn, 1995;
Meyer, 1997), there does not yet exist a generally accepted model.

In spite of the difficulties associated with the upper and lower
rings, the consensus appears to be that the GWBB model has proven
effective for SN87A (Crotts 1998).  Its success has allowed
researchers to pose basic questions about the progenitor.  In
particular, the need for an aspherical RSG wind has led a number of
authors to posit the existence of a binary companion for the
progenitor of SN87A.  Soker (1998) and Livio (1998) have argued that
binary companions are required to produce slow toroids in PNe; similar
arguments hold for SN87A (Livio 1997).  Using considerations from
stellar evolution theory, along with the need for a high equator to
pole contrast, Podsiadlowski (1992) also argued that SN87A was a
binary system.  He calculated a likely companion mass of $3$ to $6$
\msol with the progenitor/primary of $16$ \msol.  Podsiadlowski
concluded that the two stars merged before the supernova, which is
consistent with the lack of observational evidence for a companion in
the post-supernova epoch (\eg Crotts, Kunkle \& Heathcote, 1995).

In this paper, we take up the challenge posed by previous models:
whereas previous investigations have used an {\em ad hoc} description
of the aspherical RSG, we will link the shape of this slow torus to
the properties of the star itself.  Currently, the best generic
scenarios for developing asphericity in a slow wind are Common
Envelope interactions in binary stars (Livio \& Soker 1988) and the
Wind Compressed Disk (WCD) model of Bjorkman \& Cassinelli (1993). In
the Common Envelope model, a slow, dense wind which has higher mass
loss in the orbital plane of the binary system is ejected by the
primary.  While this process has been shown to be effective (Terman
\etal, 1995; Rasio \& Livio, 1996), calculations of the wind shape
require numerical models which must span a large range in both length
and time scales.  Thus, there is no simple means for linking initial
states of the binary with the shape of the ejected slow torus.  The
WCD model, which was developed for excretion disks surrounding B[e]
stars, relies on the equatorial focusing of wind streamlines from
rapidly rotating stars.  The advantage of the WCD mechanism is its
relatively simple formalism, which relates the properties of a single
star (mass, temperature, mass loss rate, rotation rate) with the
properties of the slow torus.  The WCD mechanism has been adapted to
Red Giant stars (Ignace, Bjorkman \& Cassinelli, 1996) and has already
been used to produce slow tori in LBV and PNe bipolar outflow
simulations (Garcia-Segura \etal, 1997a; Garcia-Segura \etal, 1997b).

The goal of the present paper is to utilize the WCD and GWBB
formalisms to establish the efficacy of the the combined model for
SN87A.  We then work backwards to bracket the properties of the
progenitor based on the comparison of models with observations.  In
what follows, we describe self-similar models of the nebula surrounding
SN87A, which employ a WCD model to determine the properties of the RSG
wind.  Our self-similar model is based on the work of Giuliani (1982)
and Dwarkadas, Chevalier \& Blondin (1995; hereafter DCB95). We use
observations of the rings surrounding SN87A to constrain the
parameters of the system, and to explore the implications for the
rotation rate of the progenitor.  In \S 2 we describe the WBB and WCD
models.  In \S 3 the results of the models and a comparison with
observations are presented, as well as a ``fiducial,'' or ``best-fit''
model for the nebula.  In \S 4 the implications of these results for
the progenitor are explored.  Lastly, in \S 5 we discuss our results.

\section{Theory}

When a stellar wind ``turns on,'' it expands ballistically until
enough ambient material is swept up for significant momentum to be
exchanged between the wind and the ambient medium (Koo \& McKee 1992).
A triplet of hydrodynamic discontinuities then form, defining an
``interaction region'' bounded internally (externally) by undisturbed
wind (ambient) gas.  The outer boundary is an outward-facing shock. It
accelerates, compresses, and heats the ambient material as it
propagates.  We refer to this as the {\it ambient shock}, and denote
its position as $R$.  The inner boundary is defined by an
inward-facing shock which decelerates, compresses and heats the
stellar wind.  We refer to this feature as the {\it wind shock}.  Its
position is $R_{ws}$.  A {\it contact discontinuity} (CD), $R_{cd}$,
separates the shocked wind and shocked ambient material.  In the 1
dimensional (1-D) bubble these discontinuities form a sequence in
radius: $R_{ws} < R_{cd} < R$.

The compressed gas behind either or both shocks emits strongly in
optical, UV and IR wavelengths producing a bright shell which defines
the observable ``bubble.''  In SN87A, compressed material at the
equator defines the central ring.  The dynamics of the bubble and its
emission characteristics are defined by the efficiency of post-shock
shock cooling.  Behind each shock we can define a cooling time-scale
$t_c = E_t/\dot{E_t}$, where $E_t$ is the thermal energy density of
the gas.  Radiative cooling can be expressed in terms of a cooling
function: $\dot{E}_t = C(T) = n^2\Lambda(T)$, where $n$ is the number
density of the gas, $T$ its temperature, and the function $\Lambda(T)$
is a sum over many radiative processes emitting at a variety of
wavelengths.  The bubble has cooling and dynamical time scales defined
as
\begin{eqnarray}
  t_c &=& {3  k T \over 2 n \Lambda(T)}\\
  t_d &=& {R\over V},
\end{eqnarray}
\noindent where $V$ is the speed of the ambient shock, $k$ is
Boltzmann's constant, and we have assumed (as we shall throughout this
paper) that $\gamma = 5/3$.  Comparison of $t_c$ and $t_d$ separates
WBBs into two classes: Radiative (referred to as {\em momentum
conserving}) and Adiabatic (or {\em energy conserving}).

In what follows we assume that the BSG wind speed is large enough that
the shocked wind is extremely hot, and cools inefficiently \ie $t_c
\gg t_d$. In this case the wind shock is adiabatic and hot shocked
wind material fills the expanding bipolar bubble, causing expansion
through thermal pressure, so that a {\it hot bubble} forms.  We also
assume that the shocked ambient material cools efficiently, so that
$R_{cd}\approx R$.  DCB95 showed that for thermal-pressure-supported,
or {\em energy conserving} bubbles, the expansion is self-similar for
a significant portion of its evolution.  The resulting, simplified
equations may then be solved without resorting to time-dependent
hydrodynamical calculations.

The parameters of this wind-blown bubble (WBB) model include the
description of the BSG and RSG winds.  Specifically, the wind speeds
and mass loss rates of the two winds must be supplied, as well as the
dependence of the wind speeds on the polar angle $\theta$, given by the
asymmetry functions $f(\theta)$ for the RSG wind density, and
%JEB
%$g(\theta)$ for the RSG wind speed.
$g(\theta)$ for the RSG wind speed, defined by
\begin{eqnarray}
  g(\theta) &=& \lim_{r \rightarrow \infty} v_r / v_a  ,
\label{eq:gdef} \\
  f(\theta) &=& \lim_{r \rightarrow \infty} 4\pi r^2 v_a \rho/{\dot M}_a 
.
\label{eq:fdef}
\end{eqnarray}
%JEB

\noindent In previous models of SN87A, these asymmetry functions have
been determined {\em ad hoc} (\eg Luo \& McCray, 1991; BL93). Here we
assume that the balance of wind driving and centrifugal forces focuses
the RSG wind toward the equatorial plane, creating a {\em
wind-compressed disk} (WCD: Bjorkman and Cassinelli, 1993).  A
solution of the WCD equations yields $f(\theta)$ and $g(\theta)$.  We
note again that all previous studies assumed an isotropic RSG wind
velocity, so that $g(\theta)$ is constant, whereas the WCD model
naturally proscribes a wind asphericity. In the following sections we
describe the models used for the wind blown bubble and the
environment.

\subsection{Wind-Blown Bubble Model}

The general equations describing the evolution of a thin shell have been
derived by Giuliani (1982).  In the DCB95 model, these equations are
applied to a shell of swept-up, shocked Asymptotic Giant Branch star
wind material which forms the bright optical regions of Planetary
Nebulae.  Here we apply the model to the swept-up RSG gas.  It is
assumed that the width of the shell is much smaller than its radius,
and that magnetic fields may be neglected.  The velocity and density are
averaged over the width of the shell.  We briefly sketch the
derivation of the equations here; the complete derivation may be found
in DCB95.

The continuity equation for the shell is given by
\be
  {\partial\sigma\over\partial t} = -\rho_o(v_{\perp o}-u_\perp)
	+\rho_i(v_{\perp i}-u_\perp)
	-\sigma{\partial\ln{A}\over\partial t}
	-{1\over A}{\partial\over\partial\theta}
	  \left(R\sin\theta\sigma(v_\parallel-u_\parallel)\right).
\ee
Here $\sigma$ is the internal column density of the shell, $\rho$ is
the volume density, $\bf{v}$ is the wind velocity, and $\bf{u}$ is the
shell velocity.  The independent variables are the spherical radius
$r$, the spherical polar angle $\theta$, and the time $t$, and
$R(\theta, t)$ is the radius of the shell.  The quantities inside the
bubble are labeled with the subscript ``o'', while those outside the
bubble bear the subscript ``i,'' and quantities which are averaged
over the width of the shell have no subscript.  The velocity vectors
have been decomposed into components perpendicular ({\em e.g.}
$u_\perp$) and parallel ({\em e.g.}  $u_\parallel$) to the shell.
The quantity $A$, used for simplicity of notation, is defined by
\be
  A\equiv\left({r^2\sin\theta\over\cos\xi}\right)_R,
\ee
where $\xi$ is the angle between the radius vector and the local
normal to the shell surface.  The defining equation for $\xi$ is
\be
  \tan\xi\equiv-{1\over R}{\partial R\over\partial\theta}.
\ee
The first two terms on the right-hand side (rhs) of equation (6)
represent the flux of mass through the inner and outer surfaces of the
shell.  The third term represents the change in surface density
due to the change in surface area of the bubble, while the final
term accounts for flow within the shell.

The perpendicular and parallel components, respectively, of the
momentum equation are given by
\begin{eqnarray}
  \sigma\left({\partial \vpe\over\partial t}
	-\vpa{\partial\xi\over\partial t}\right)
  &=& -\left[\rho_o(\vpeo-\upe)(\vpeo-\vpe)\right]\nonumber\\
  & & +\left[\rho_i(\vpei-\upe)(\vpei-\vpe)\right]+P_i\nonumber\\
  & & -(\vpa-\upa){\sigma\cos\xi\over R}
	\left[{\partial\vpe\over\partial\theta}
	-\vpa\left(1+{\partial\xi\over\partial\theta}\right)\right],
\end{eqnarray}
and
\begin{eqnarray}
  \sigma\left({\partial \vpa\over\partial t}
	-\vpe{\partial\xi\over\partial t}\right)
  &=& -\left[\rho_o(\vpeo-\upe)(\vpao-\vpa)\right]\nonumber\\
  & & +\left[\rho_i(\vpei-\upe)(\vpai-\vpa)\right]\nonumber\\%+P_i
  & & -(\vpa-\upa){\sigma\cos\xi\over R}
	\left[{\partial\vpa\over\partial\theta}
	+\vpe\left(1+{\partial\xi\over\partial\theta}\right)\right],
\end{eqnarray}
where $P_i$ is the thermal pressure causing bubble expansion, and we
have assumed that $P_i\gg P_o$.  In equations (8) and (9), the first
two terms on the rhs represent the ram pressure balance.  In equation
(8) the third term represents the thermal pressure.  The derivatives
of $\vpa$ and $\vpe$ with respect to $\theta$ in the final terms of
equations (7) and (8) account for the ram pressure balance within the
shell, and the term proportional to $(1+\partial\xi/\partial\theta)$
is due to the centrifugal force.

The order of these equations may be reduced if one assumes that the
flow is self-similar -- {\em i.e.}, that the expansion velocity is
proportional to the spherical radius $r$.  With this assumption, the
radius $R$ may be written in terms of a self-similar radius $L$, using
the expansion velocity at the pole, $v_p$: $R(\theta, t) =
v_ptL(\theta)$.  For a RSG wind with constant mass flux $\dot{M}_a$
and radial velocity $v_a$, the density outside the bubble may be
written
\begin{equation}
  \rho_o={\dot{M}_af(\theta)\over 4\pi v_aR^{2}},
\end{equation}
where $f(\theta)$ is the asymmetry function, which is identically one
for an isotropic wind.  The self-similar surface density $S(\theta)$
is defined by
\begin{equation}
  \sigma = {\dot{M}_aS(\theta)\over4\pi v_av_pt}.
\end{equation}
The parallel component of the gas velocity within the shell is given by
\begin{equation}
  \vpa=v_p U(\theta).
\end{equation}
The parallel and perpendicular components of the shell velocity are
defined by
\begin{equation}
  \upa=-{\partial R\over\partial t}\sin\xi,\qquad
  \upe= {\partial R\over\partial t}\cos\xi,
\end{equation}
so that
\begin{equation}
  \upa=-v_p\sin\xi L(\theta), \qquad \upe=v_p\cos\xi L(\theta).
\end{equation}
Because the BSG wind undergoes a strong shock, $\upe\approx\vpe$.  The
velocities outside the bubble are
\begin{equation}
  \vpeo=v_ag(\theta)\cos\xi,\qquad \vpao=-v_ag(\theta)\sin\xi,
\end{equation}
where $g(\theta)$ is the asymmetry of the RSG wind velocity.  Because
the RSG wind is assumed to have a constant mass flux, $f(\theta)=
1/g(\theta)$.  We assume without loss of generality that $f(0)=1$, and
that $L(0)=1$.  Finally, the pressure inside the bubble is
proportional to $\rho v^2$.  Given that the mass flux is constant,
$\rho\sim r^{-2}$.  Since $v\sim r/t$, we expect $P_i\sim t^{-2}$.
Thus, we let $P_i=F/t^2$, where $F$ is a constant.

With these substitutions, equations (5) and (7)-(9)  become
\begin{eqnarray}
  L^\prime &=& -L\tan\xi,\\
  \xi^\prime &=& {L\over S}(U+L\sin\xi)^{-2}\Bigl[f(\theta)\cos\xi\nonumber\\
    & & \times\left({\lambda_v\over L}-1\right)^2
      -\sec\xi(\lambda-1)^2\Bigr]-1,\\
  U^\prime &=& {f(\theta)\over S}\left({\lambda_v\over L}-1\right)
    \left(U+\lambda_v\sin\xi\over U+L\sin\xi\right)
    -L\cos\xi(1+\xi^\prime),\\
  S^\prime &=& S(\tan\xi-\cot\theta)-f(\theta)(U+L\sin\xi)^{-2}\nonumber\\
    & & \times\left({\lambda_v\over L}-1\right)
     (2U + (L+\lambda_v)\sin\xi),
\end{eqnarray}
where we have defined $\lambda\equiv v_a/v_p$, and $\lambda_v\equiv
\lambda g(\theta)$.

At the poles
there is no flow within the shell, so that $\xi(0)=U(0)=0$.  In the
limit $\theta\to0$, the self-similar form of the perpendicular
component of the momentum equation (8) yields the equation
\begin{equation}
  F = {\dot{M}\over4\pi v_a}(\lambda-1)^2.
\end{equation}
This has been used to eliminate $F$ in equation (15).

The value of $v_p$ is determined using the equation for bubble energy
conservation.  The rate of change of internal energy $E$ is given by
the work done by the fast wind, $l$, minus the work done to expand
the bubble, $dW/dt$:
\begin{equation}
  {dE\over dt}=l-{dW\over dt}.
\end{equation}
As is shown in DCB95, this yields the constraint on $v_p$ of
\begin{equation}
  \chi(v_a-v_p)+v_p^{-1/2}\left({v_f^2v_a\dot{M}_f\over3\dot{M}_a}\right)^{1/2}
  =0,
\end{equation}
where $\dot{M}_f$ is the BSG mass loss rate, and $\chi$ is the ratio
of the volume of the bubble to the volume of a sphere with the same
polar radius.

The system of equations of solved as follows: A value of $\lambda$ is
assumed.  Equations (16)-(19) are then solved numerically to determine
the self-similar functions $L(\theta)$, $S(\theta)$, $U(\theta)$ and
$\xi(\theta)$.  Given $L(\theta)$ it is possible to calculate
numerically $\chi$.  Equation (22) then yields a new value of $v_p$,
and thus a new value of $\lambda$.  Iteration of this procedure may be
used to find self-consistent energy-conserving models of the nebula
surrounding SN87A.

The four variables $v_a$, $v_f$, $\dot{M}_f$ and $\dot{M}_a$ are
parameters of the system which are varied to match observations of
SN87A.  The RSG wind density asymmetry function $f(\theta)$ is
determined using the wind-compressed disk model, described in \S 2.2,
and the velocity asymmetry function $g(\theta)$ is, as mentioned
above, given by its inverse.

Finally, we note that an implementation of the fifth-order Runge-Kutta
algorithm was used to solve the WBB equations (16)-(19).  A Newton's
method was used to solve the additional constraint of equation (22).

\subsection{Wind-Compressed Disk Model}

%JEB
The Wind-Compressed Disk Model (Bjorkman \& Cassinelli 1993, hereafter
BC) was developed to determine the structure of a rotating stellar
wind.  Their model applies whenever the external forces driving the
wind are central (\ie, radial) forces.  BC found that, in the
supersonic portion of the flow, the wind streamlines originating at
high latitudes converge toward the equator.  The reason for this
behavior may be understood as a simple consequence of orbital
dynamics.  As the fluid leaves the surface of the star, it tends to
orbit around the star until it is accelerated radially outward.  If
the initial outward acceleration is small compared to the rotation
rate of the star, then the material orbits far enough around the star
that it crosses the equator, where it meets material from the opposite
hemisphere of the star.  Since the flow velocities perpendicular to
equator are supersonic, a pair of shocks form above and below the
equator, and the shock compression of the material entering the
equatorial region creates a dense disk.  Thus we see that whether or
not a disk forms depends on the rotation rate of the star, $\Omega =
V_{\rm rot} / V_{\rm crit} = \Omega_* / \Omega_K$, where $V_{\rm
rot}$ and $V_{\rm crit}$ are the rotation and break-up speeds, and
$\Omega_*$ and $\Omega_K$ are the stellar and Keplerian rotation
rates.  If the star rotates faster than a threshold value, $\Omega >
\Omega_{\rm th}$, a dense, shock-compressed disk forms, with $p \sim
100$. Otherwise, there is only a mild density compression in the
equator (typically $p<3$--10).  BC originally applied this model to
investigate if the disks around Be stars could be produced by the WCD
mechanism.  The WCD model was later applied to several other classes
of stars, including AGB stars, by Ignace, Cassinelli, \& Bjorkman
(1996, hereafter ICB).

To model the RSG wind, we follow ICB to obtain the flow velocity and
density of the wind.  ICB assume that the wind velocity of AGB stars
can be described by the usual $\beta$-law,
\begin{equation}
  v_r(r,\theta) = v_0 + [v_\infty(\theta_0) - v_0]\left(1 -
                                             R_{\ast}/r \right)^{\beta}  ,
\label{eq:betalaw}
\end{equation}
where the initial velocity $v_0$ equals the sound speed, and $v_\infty$ is the
wind terminal speed.  The value of the velocity law exponent, $\beta$,
determines the acceleration of the wind.  ICB assume that the winds of AGB
stars
are slowly accelerating, so they adopt $\beta=3$ in their model.

The wind density is determined by the continuity equation, which yields
\begin{equation}
  \rho(r,\theta) = {{{\dot M}(\theta_0)}\over{4 \pi r^2 v_r (d\mu / d\mu_0)}}
 ,
\label{eq:wcdrho}
\end{equation}
where $\mu=\cos \theta$, $\mu_0=\cos \theta_0$, and $\theta_0$ is the
initial co-latitude (at the stellar surface) of the streamline passing
through the point ($r,\theta$).  Note that the factor $(d\mu /
d\mu_0)$, given by equation (B3) of ICB, is the solid angle of the
streamtube; this change in cross-sectional area produces the
equatorial wind compression and is primarily responsible for the
increased density near the equator.

To evaluate the density, we must first determine $\theta_0$ for the
given location ($r,\theta$) by solving numerically equation (B2) of
ICB.  However for our wind-blown bubble model, we are only interested
in distances that are quite far away from the star.  In the limit $r
\rightarrow \infty$, equation (A6) of ICB for the azimuthal deflection
of the streamline becomes
\begin{equation}
\phi^\prime_{\rm max}(\theta_0) = {{V_{\rm rot} \sin \theta_0}\over{\beta v_0}}
                 \left({{v_0}\over{v_\infty(\theta_0)-v_0}}\right)^{1/\beta}
   B_{y_{\rm max}(\theta_0)}\left({{1}\over{\beta}},1-{{1}\over{\beta}}\right)
     ,
\label{eq:phimax}
\end{equation}
where $B$ is the incomplete gamma function with the argument $y_{\rm
max}(\theta_0) = 1 - v_0/v_\infty(\theta_0)$.  In the large-$r$ limit,
$\theta_0$ is given by the solution to the equation
\begin{equation}
    \cos \theta_0 \cos \phi^\prime_{\rm max}(\theta_0) = \cos \theta .
\label{eq:theta0}
\end{equation}
With this value for $\theta_0$, we then evaluate the wind compression factor
$(d\mu / d\mu_0)$, given by equation (B3) of ICB, as well as the streamline
mass
loss rate and terminal speed, which are given by
\begin{eqnarray}
    {\dot M}(\theta_0) &=& {\dot M}_a \left(1-\Omega \sin \theta_0\right)^{\xi}
      ,
\label{eq:wcdMdot} \\
    v_{\infty}(\theta_0) &=& v_a \left(1-\Omega \sin \theta_0\right)^{\gamma}
      ,
\label{eq:wcdvinf}
\end{eqnarray}
with $\gamma=0.35$ and $\xi=-0.43$ (see BC).

Substituting equations (\ref{eq:wcdrho}), (\ref{eq:wcdMdot}), and
(\ref{eq:wcdvinf}) into equation (\ref{eq:gdef}) and (\ref{eq:fdef}),
we find the wind asymmetry functions
\begin{eqnarray}
  g(\theta) &=& \left(1-\Omega \sin \theta_0\right)^{\gamma}  ,
\label{eq:gwind} \\
  f(\theta) &=& \left(1-\Omega \sin \theta_0\right)^{\xi-\gamma}
                (d\mu / d\mu_0)^{-1}
                 .
\label{eq:fwind}
\end{eqnarray}

These equations are valid for the wind.  However, if the stellar
rotation rate is above the disk formation threshold ($\Omega >
\Omega_{\rm th}$), then we need the disk properties as well.
Unfortunately, the WCD model {\em per se} cannot determine the disk
density and velocities, so we make a simple estimate based on the
numerical hydrodynamics simulations by Owocki, Cranmer \& Blondin
(1994).  They found that the disk has an opening angle $\Delta
\theta_d \approx 3^\circ$, which agrees with observations of Be stars
(Wood, Bjorkman \& Bjorkman 1997), and that the disk terminal speed
$v_d \sim {\hbox{0.2--0.3}} v_a$.  However based on observations of
the disks around Be and B[e] stars (Waters \etal 1988, Zickgraf \etal
1996), we adopt a slightly lower value $v_d = 0.1 v_a$.  In addition
to the geometry and flow speeds, Owocki \etal found that there is a
stagnation point in the disk.  Interior to the stagnation point, the
disk material falls back onto the star, while exterior to the
stagnation point the disk material flows outward.  At low rotation
rates, this recirculation is not dominant, so for simplicity we assume
that all the wind material entering the disk flows outward.  Using the
continuity equation, the disk density may be estimated by
\begin{equation}
    \rho_d = {{{\dot M}_d}\over{2 \pi r^2 \sin \Delta \theta_d v_d}}  .
\label{eq:diskrho}
\end{equation}
The mass loss rate entering the disk,
\begin{equation}
    {\dot M}_d = {1\over2} \int_0^{\mu_0^d} {\dot M}(\theta_0)\,d\mu_0  ,
\label{eq:Mdotdisk}
\end{equation}
is determined by $\mu_0^d =\cos \theta_0^d$, the minimum co-latitude
of all streamlines that enter the disk (see Fig.~10 of BC), which is
given by the solution to
\begin{equation}
    \phi^\prime_{\rm max}(\theta_0^d) = \pi/2  .
\end{equation}

Using the disk terminal speed, $0.1 v_a$, and equation
(\ref{eq:diskrho}), we find that the velocity and density asymmetry
functions for the disk are
\begin{eqnarray}
    g_d &=& 0.1  ,
\label{eq:gdisk} \\
    f_d &=& {{\int_0^{\mu_0^d} (1-\Omega \sin\theta_0)^{\xi}\,d\mu_0} \over
           {g_d \sin \Delta \theta_d }}  .
\label{eq:fdisk}
\end{eqnarray}
These equations are to be used only when $\Omega > \Omega_{\rm th}$ {\it and }
$|\theta - \pi/2| < \Delta \theta_d$; otherwise, one should use equations
(\ref{eq:gwind}) and (\ref{eq:fwind}).

\section{Results}

As was discussed in \S 1, the accepted WBB model for SN87A treats the
inner ring as the ``waist'' of the bipolar outflow and assumes that
the outer rings lie somewhere on the surface of the lobes or represent
lobe edges.  In what follows we do not attempt to explain the outer
rings, and primarily use the inner ring to constrain the parameters of
our models.

We use a number of observed features of SN87A for comparison with the
models: The radius of the inner ring, $\varpi \sim6.3\times10^{17}$
cm, sets the length-scale of our solutions.  The density of the inner
ring has been found to be at least $\sim(1-2)\times10^3$ \cc
~(Lundqvist \& Sonneborn, 1997).  Based on optical images, the width
of the inner ring, $\delta r$, is approximately one tenth of its
radius.  We use this to convert from column density to volume density:
$\rho_\varpi = \sigma/\delta r$.  The radial velocity of the inner
ring is $v_\varpi \approx 8.3$ \kms (Meaburn \etal 1995).  The
quantities $v_\varpi$ and $\rho_\varpi$ are the primary observations
used to constrain the parameters of our models.  We note that the
polar expansion velocity may be given in terms of the velocity
asymmetry function $g(\theta)$ as $v_p=v_\varpi f(\pi/2)$.  In
addition to $v_\varpi$ and $\rho_\varpi$, we have calculated the ratio
of the cylindrical radius of the lobes at their widest points to the
radius of the inner ring (which we denote $R_o^\prime$).  This is
compared with the ratio of the cylindrical radius of the outer rings
to that of the inner rings (which we denote $R_o$). The observed value
of $R_o$ is $\sim 2$.

\subsection{Fiducial Models}

As described in \S 2.2, the angular dependence of the RSG wind depends
on the ratio $\Omega$ of the progenitor's rotation rate to the breakup
rotation rate. The equator-to-pole density contrast $p$ of the RSG
wind is therefore proportional to $\Omega$.  In what follows, we
assume that the RSG has a radius of 740 R$_\odot$, a mass of 20
M$_\odot$, and an effective temperature of 3300 K.  Given these
values, the WCD model shows that when $\Omega \sim 0.3$, the
equator-to-pole density contrast becomes significant, $p\sim25$, even
if a shock-bounded ``disk'' does not form.  For $\Omega\gtrsim0.3$, a
thin excretion disk forms with a very high density contrast of
$p\gtrsim700$. Upon driving a BSG wind into these environments we find
we can produce a bipolar wind-blown bubble consistent with the nebula
around SN87A.  In particular, we can recover the observed properties
of the inner ring with two models, one with and one without a disk.
The shape, expansion velocity and surface densities of these models
are shown in Figs. 1, 2, and their density asymmetry functions
$f(\theta)$ are shown in Fig. 3.  The initial parameters and the ring
properties for the two models are given in Table 1. In both cases the
expansion speed of the ring is equal to $\sim8.3$ \kms.  (We have also
created a model with a ring expansion speed of $\sim10$\kms, in
accordance with the observations of Crotts {\em et al.}, 1995, and
find that it requires, for the model with the WCD, an $\Omega$ lower
by $0.02$; This does not significantly affect the results given by the
Binary Merger Model of \S 4.)  The ring density is within a factor of
two of observed estimates (Table 1).

It is important to note that we can produce successful models without
resorting to anomalously low values for the speed of the RSG wind at
the poles.  As was discussed in \S 1, previous studies of SN87A have
been forced to use velocities as low as $v_a\sim5$ \kms~ to recover
the low ring expansion speed, $v_\varpi$.  The hydrodynamics of Wind
Compression naturally reduces $v_a$ at the equator, since mass
conservation requires that the density is inversely proportional to
the velocity.  The reduced wind speeds at the equator allow us to use
larger, more canonical values, of the RSG wind speed.

The models with and without a disk differ most notably in the ratio
$R_o$.  Assuming that the outer rings lie at the widest point on the
lobes, the observed value is $R_o\sim2$.  The model with the disk has
$R_o=4.9$, whereas for the model without the disk, $R_o=2.6$.  The
difference can be explained as follows: When a disk forms, its high
density severely inhibits expansion of the ambient shock at the
equator.  A discontinuity in the shell radius $L(\theta)$ occurs just
above (below) the equator in these models.  The BSG wind is able to
expand rapidly along the the top and bottom edges of the disk
producing a strongly ``wasp-waisted,'' or pinched bubble. It is
possible that such configurations are an artifact of our calculations.
Numerical simulations have demonstrated that the lobes of energy
conserving bubbles can experience considerable shaping due non-linear
effects associated with the wind shock.  If the wind shock is not
spherical but assumes a prolate geometry, then post-shock BSG gas is
focused towards the poles (Frank \& Mellema, 1996).  The higher ram
pressures associated with this ``shock focusing'' will tend to
elongate the bubble, reducing $R_o$ (BL94).  Frank \& Mellema (1994)
found that higher values of $p$ produce stronger shock focusing. These
effects will not be seen with a self-similar method which assumes the
shell is driven by an isobaric ``hot bubble''. An isobaric bubble can
not be produced with an aspherical wind shock.

Figs. 1 and 2 and the ring properties given in Table 1 demonstrate the
success of the unified GWBB and WCD models for SN87A.  This represents
one of the principal conclusions of our study. We take the cases
represented in Figs. 1 and 2 as our fiducial models.

\begin{table}[tb]
\centering
\begin{tabular}{ccccc|ccc}
  \hline\hline
  $\Omega$ & $v_a$ & $\dot{M}_a$ &
  $v_f$ & $\dot{M}_f$ &
  $v_\varpi$ & $n_{\rm eq}$ &
  $R_o$ \\
   & km s$^{-1}$ & $10^{-5} $M$_\odot$\ yr$^{-1}$ &
     km s$^{-1}$ & $10^{-7} $M$_\odot$\ yr$^{-1}$ & km s$^{-1}$ &
     cm$^{-3}$ & \\
  \hline
  \multicolumn{5}{c|}{Fiducial Model 1 : Wind Compressed Zone} & & & \\
  \hline
  0.3 & 20 & 2 & 400 & 1 & 8.30 & 0.87 & 2.6 \\
  \hline
  \multicolumn{5}{c|}{Fiducial Model 1 : Wind Compressed Disk} & & & \\
  \hline
  0.36 & 20 & 1 & 600 & 3 & 8.36 & 7.0 & 4.9 \\
  \hline
  \multicolumn{5}{c|}{variation of parameters:} & & & \\
  \hline
  0.33 & 20 & 2 & 400 & 1 & 7.02  & 1.9  & 3.2 \\
  0.3  & 25 & 2 & 400 & 1 & 8.23  & 0.73 & 2.7 \\
  0.3  & 20 & 1 & 400 & 1 & 10.5  & 0.47 & 2.5 \\
  0.3  & 20 & 2 & 600 & 1 & 10.7  & 0.94 & 2.4 \\
  0.3  & 20 & 2 & 400 & 3 & 11.7  & 0.97 & 2.4 \\
  \hline\hline
\end{tabular}
\caption{Dependence of the properties of the models (inner ring
  expansion speed $v_\varpi$, inner ring number density $n_\varpi$,
  and ratio of width of outer to inner rings, $R_o$) on the input
  parameters ($\Omega$ ratio of RSG rotation rate to breakup rotation
  rate, RSG wind speed $v_a$ and mass loss rate $\dot{M}_a$, BSG wind
  speed $v_f$ and mass loss rate $\dot{M}_f$).  }
\end{table}

% What are the basic features of the fiducial model?
% 1. it posesses a bipolar bubble
% 2. The observed ring expansion speed may be matched without
%    resorting to anomolously low RSG wind speeds
% 3. The number density is correct.

\subsection{Variation of Parameters}

We have varied the five input parameters to determine (1) the effect
each has on the properties of the solutions, and (2) the degree to
which the input parameters constrain the solutions.  In particular, we
are interested in the range of $\Omega$ necessary to match
observations, given that the RSG and BSG wind parameters are within
the canonical range of values (\eg $v_a\sim20$ \kms, $400\lesssim
v_f\lesssim 600$ \kms, $\dot{M}_a\sim10^{-5}$ \mdot, $\dot{M}_f\sim
10^{-7}$ \mdot).  We find we can bracket the appropriate values of
$\Omega$. Our results show that solutions outside the range
$0.3\lesssim\Omega_c\lesssim0.36$ do not produce reasonable results for
the bubble parameters.  We therefore confine our discussions to cases
within that range.

As mentioned above, we find two models which provide agreement with
the observed ring expansion speed: one with a disk, and one without.
As $\Omega$ increases, so does the pole-to-equator density contrast
(and thus $R_o$).  This results in higher equatorial densities, and
thus smaller central ring expansion speeds.  For a critical value of
$\Omega=\Omega_c$ such that $0.3\lesssim\Omega_c\lesssim0.36$, the
wind asymmetry function develops a discontinuity with respect to
$\theta$, representing the presence of a disk.  Models with a disk
have considerably higher equatorial density, and must have
correspondingly higher values of the BSG speed and mass loss rate if
the same ring expansion speed is to be obtained (see Table 1).

Increasing $v_f$ or $\dot{M}_f$ results in a greater BSG wind momentum
flux.  This causes the central ring to expand at a higher rate, so
$v_\varpi$ is higher.  This enables the bubble to sweep up more
 of the RSG wind per unit time, resulting in a larger central
ring density.

Models with a higher value of $v_a$ possess a weaker RSG shock,
resulting in lower equatorial expansion speeds and lower equatorial
densities.  When the RSG mass loss rate is decreased, $v_\varpi$
increases, due to the smaller inertia of the RSG wind.  Also, a smaller
$\dot{M}_a$ means less mass is swept-up by the expanding bubble; thus,
the ring density is lower.

Table 1 shows that the combined WBB/WCD model can be strongly
constrained by input parameters and observations.

\section{Binary Merger Model}

As was seen in the previous section, for reasonable wind speeds and
mass-loss rates, the RSG must have been rotating at least at $\sim0.3$
of the breakup rotation rate.  This is higher than can be expected for
a single star (Fukuda, 1982; {\it cf.} Heger \& Langer, 1998).  Thus,
our results imply the presence of a binary companion in order to
spin-up the primary (Soker, 1998).  In this section we present a
simple model for a binary merger, and find the companion mass needed
to achieve the rotation rates found in \S 3.2.  Since we assume the
companion to be of a much lower mass than the primary, it is expected
that once the primary fills its Roche lobe, a dynamical mass transfer
process will ensue (\eg Rasio \& Livio, 1996) with the subsequent
spiraling-in of the companion, inside the common envelope.

In order to obtain a simple lower limit
estimate for the ratio of the mass of the
companion to the mass of the primary, $q\equiv M_p/M_c$, we assume
that in the process of coalescence, the companion mass is added
to the core of the primary, while all of its angular momentum serves to
spin-up the primary's envelope (see Fig. 4).
The initial angular momentum of the system is given by
\begin{equation}
  l_i = \left[{G(M_pM_c)^2\over M_p+M_c}a\right]^{1/2},
\end{equation}
where $a$ is the distance between the center of masses of the primary
and companion.
Given that the angular momentum after coalescence is primarily that of
the envelope, the final angular momentum is
\begin{equation}
  l_f = \xi M_e R_e^2\Omega_r,
\end{equation}
where $\Omega_r\equiv\Omega\Omega_K$ is the rotation rate (and
$\Omega_K$ the breakup rotation rate), $\xi$ is the square of the
gyration radius,
$R_e$ is the radius, and $M_e=M_{e0}=M_{e1}$ is the envelope mass,
which is assumed to remain the same during coalescence.
Setting $l_i = l_f$ and
simplifying, we find
\begin{equation}
  M_e^{-1}M_cM_p(M_p+M_c)^{-1/2}(M_e+M_{c1})^{-1/2}=\xi
  \left({R_e\over a}\right)^{1/2}\Omega.
\end{equation}
We assume that the final radius is approximately equal to the ``volume
equivalent'' radius of the Roche lobe of the primary, $R_V=a\tilde{R}_V(q)$,
where $\tilde{R}_V$ may be approximated by the function
\begin{equation}
  \tilde{R}_V = {0.49q^{2/3}\over0.6q^{2/3}+\ln(1+q^{1/3})}
\end{equation}
(Eggleton, 1983), valid for all $q$.  Using the assumption that
$M_c\ll M_p$, equation (38) may be further simplified, as
\begin{equation}
  (1+q)^{-1}\approx\xi\tilde{R}^{1/2}_V(q)\Omega.
\label{coreq0}
\end{equation}

If we assume that the total mass is given by $M_c+M_p=20M_\odot$ [so
that $q = M_p/M_c=(20M_\odot-M_c)/M_c=20/M_c-1$] (Podsiadlowski,
1992), we may solve for the lower limit on the companion mass:
\begin{equation}
  M_c\approx20M_\odot\xi\Omega\tilde{R}_V(q).
\label{coreq1}
\end{equation}
For a giant star, $\xi\approx0.1$ (\eg Pringle 1974).  Equation
\ref{coreq1} is a nonlinear equation for $M_c$ which may be solved
numerically.  For these values of ($M_c+M_p$) and $\xi$ and for
$\Omega\lesssim0.5$, $\tilde{R_V}$ is close enough to constant that
$M_c$ is approximately given by the linear function $M_c\approx
1.6\Omega$.  As a result, for $\Omega=0.3$, the minimum companion mass
is $\sim0.5$ M$_\odot$.  For the model with a disk, the minimum
companion mass is predicted to be slightly higher, $M_c\approx0.6$
M$_\odot$.

We must also consider the fact that material may also be ejected in
the CE phase solely because of the the transfer of orbital energy into
the envelope.  As others have noted this is indeed a second possible
mechanism for the development of the ring.  The difficulty with using
such a scenario to link the shape of the nebula with the progenitor is
that no simple way exits to connect the properties of the binary with
the shape of the ejected slow wind and hence with the eventual shape
of the bipolar nebula.  Calculation of the ejection process requires
expensive 3-D calculations and these only provide information about
relatively small scales ($R < 10^{14}$ cm, Sanquist \ea 1998).  Thus
while envelope ejection is alternative possibility which should be
explored the advantage of the WCD model is that one can make definite
statements about its consequences for the nebula via purely analytical
methods.
%It is possible that some mass is ejected in the early
%stages of the CE before the WCD begins. In fact, Soker (1998)
%proposed such a model to explain the presence of the two outer rings.

The dividing line between the WCD and envelope ejection models is
determined by $\alpha_CE$, the ratio of the envelope binding energy to
the binary orbital energy which can be expressed as (de Kool 1990);

\begin{equation}
\alpha_{CE} = {E_{env} \over \Delta E_{orb}} 
     =  {4 a_f M_p M_c \over a_i M_c M_e - a_f M_p M_c}
\end{equation}

\noindent where $a_i$ and $a_f$ are the initial and final orbital
separations.  If $\alpha_{CE} > 1$ then the entire envelope will not
be lost in the CE phase.  Consideration of this equation for
parameters used in our models shows that if much of the orbital energy
is deposited at radii $> .01 R_e$ then $\alpha_{CE} > 1$ and we can
expect the WCD to create a equatorial disk.  It is possible that some
mass is ejected in the early stages of the CE before the WCD
begins. In fact, Soker (1998) recently proposed such a model to
explain the presence of the two outer rings.

We note also Podsiadlowski's (1992) suggestion  that a binary merger is
responsible for the transition from the RSG to BSG.  While our model is
independent of the Podsiadlowski scenario it useful to compare the
time scale for the two processes.  Podsiadlowski quotes time scales for
the merger process of a few $10^3$ years. An upper limit on the time scale
for the life of WCD is given simply by the dynamical time scale for the
equatorial slow wind material,

\begin{equation}
\tau_{WCD} \le \tau_{dyn} = {R_{ring} \over V_a}
		      ~ {10^{17} {\rm cm} \over 5 {\rm ~km~s}^{-1}} 
			\approx 10^4
\end{equation}

The dense torus resulting from the WCD does not, however, have to be as
large as the ring is currently. It could have had a much smaller
spatial extent ($R \approx 10^{16}$) and still produce a nebula of the
right size.  Dwarkadas \& Balick (1998), have shown that even a very
small ($R \approx 10^{14}$) dense ring can hydrodynamically shape a
bipolar nebula.   Thus the time scales needed to create the BSG and the
WCD appear close.

We note also from Iben \& Livio (1993) that $\gamma_{CE}$, the ratio
of the spin up time scale to the orbital decay time scale has the
following proportionality

\begin{equation}
\gamma_{CE} = {\tau_{spin} \over \tau_{decay}} 
                       \propto {\hat{\rho} \over \bar{\rho}},
\end{equation}
where $\hat{\rho}$ is the local density the secondary star experiences
as it spirals inward and $\bar{\rho}$ is the (considerably larger)
average density for the RSG. Thus for an extended star like a RSG,
$\gamma_{CE} < 1$, and the envelope will be quickly spun up allowing
the WCD mechanism to operate.

\section{Discussion and Conclusions}

We have developed a model for the bipolar nebula surrounding Supernova
1987A.  As in previous investigations, we invoke the interaction
between a fast isotropic Blue Supergiant Wind (BSG) and a slow
aspherical (toroidal) Red Supergiant Wind.  The novel aspect of our
model is the the use of the Wind Compressed Disk (WCD) model to
determine the geometrical properties of the RSG wind.  Wind
compression occurs when a star rotates fast enough to deflect wind
streamlines towards the equator.  The increased density and reduced
velocity in the equatorial-zone RSG wind provides the constraint which
shapes the final wind blown bubble (WBB).  The degree of wind
compression depends on the characteristics of the star. Thus our model
allows the observable nebular properties to be directly linked to the
unobservable properties of SN87A's progenitor.

Our results show that the combined WCD/WBB model can recover the
observed size, speed, density and gross morphology of SN87A's
circumstellar nebula.  Wind compression allows our models to recover
the low expansion speeds of the ring without resorting to anomalously
low values of the RSG wind speed.  In addition, we have shown that our
models are sensitive to initial conditions allowing us to bracket the
properties of the progenitor.  In particular our models predict that
the rotation rate of the progenitor must have been a significant
fraction of the critical speed for break-up, $\Omega \gtrsim 0.3
\Omega_K$.  Since this is too large to be expected for a single RSG we
infer that SN87A's progenitor was probably spun-up by a companion.
Since no companion is visible now, we have developed a simple model
for a binary merger.  This model, along with the results of the
WCD/WBB calculations, allows us to predict that SN87A had a companion
with a mass of $ M_c \gtrsim 0.5$ \msol.

Our value for $M_c$ clearly represents a lower limit. In addition to
the fact that we have assumed an angular momentum deposition with
100\% efficiency into the envelope, other effects also need to be
considered. The shape of the bubble and speed of the ring are
determined by the density contrast in the RSG wind.  The formulation
of the WCD model relies on an approximate method for tracing RSG wind
streamlines.  This ignores pressure effects.  At large distances from
the star it is likely that the wind compressed disk zone would
experience some re-expansion which would weaken the density contrast.
It is noteworthy that Garcia-Segura, Langer \& MacLow (1997a) in their
numerical WCD/WBB models for $\eta$ Carina found extreme stellar
rotation rates $\Omega \sim 0.9 \Omega_K$ were required to recover the
correct morphology. The rapid evolution and high stellar temperature
$T_*$ associated with $\eta$ Car make it quite different from SN87A.
However the Garcia-Segura models indicate that time-dependent
hydrodynamic effects may change the properties of the disk. Thus, for
SN87A, we expect larger values of $\Omega$ would be required to
produce high values of $p$ and the correct inner-ring expansion speed.
As equation (41) demonstrates, larger values of $\Omega$ imply higher
companion masses.

In addition, we should note that the coalescence process is likely to
produce some mass-loss which was not included in the calculation. A
more accurate estimate of $M_c$ will therefore require both a complete
common envelope evolution calculation as well as the use of numerical
simulations of the bubble evolution.

In spite of these uncertainties, our models provide a link between the
properties of the circumstellar nebula and the progenitor star.  Our
results strongly indicate that SN1978A had a binary companion which
merged with the primary before the supernova explosion.  A companion
with a mass $M_c\sim3~-~6$\msol (as suggested by stellar evolution
calculations; \eg Podsiadlowski 1992) is perfectly consistent with the
dynamical considerations in the present work.

\acknowledgements

We would like to thank Arlin Crotts, Larry Helfer and Peter Lundqvist
for helpful discussions. Support for AF and TC was provided at the
University of Rochester by NSF grant AST-9702484 and the Laboratory for
Laser Energetics. ML acknowledges support from NASA Grant NAG5-6857.

%-------------------------------------------------------------------------

\newpage
\begin{figure}[tbp]
\epsfxsize=4.5in
\centerline{\epsfbox{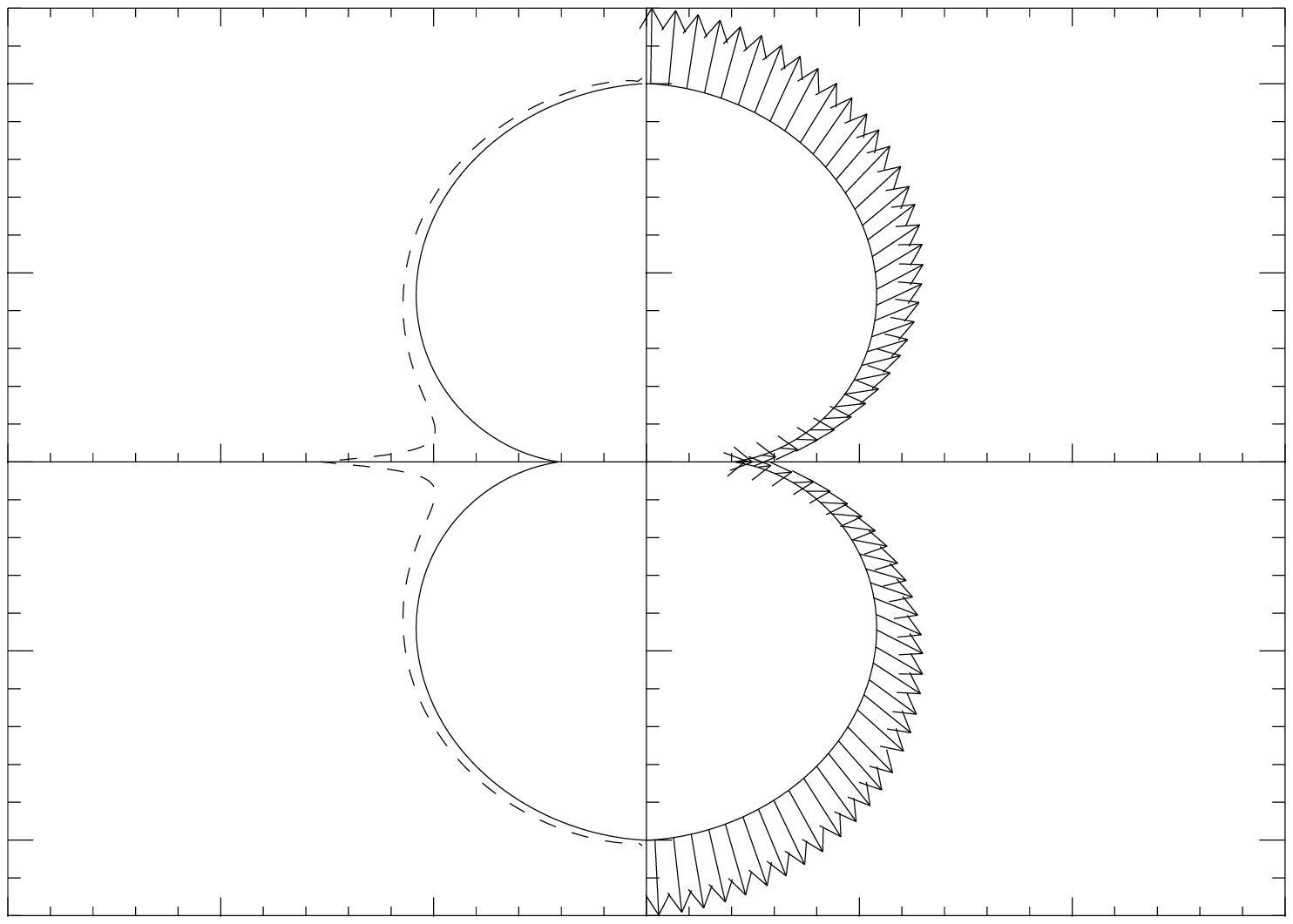}}
\figcaption{The fiducial nebula model having no disk.  The radius of the
solid line shows the shape of the bubble, and is proportional to its
expansion speed.  The arrows indicate the magnitude and direction of
the flow in the shell.  The distance between the dashed line and the
solid line is proportional to the surface density of the bubble.}
\end{figure}
\begin{figure}[tbp]
\epsfxsize=4.5in
\centerline{\epsfbox{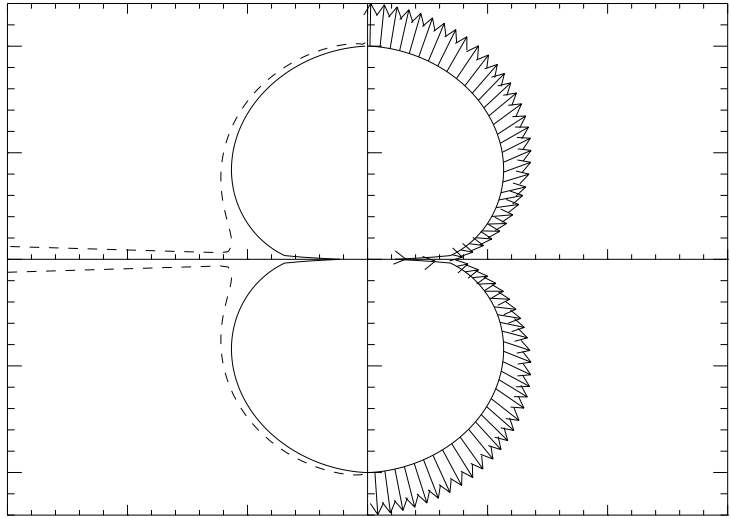}}
\figcaption{The fiducial nebula model with a disk.  As in Fig. 1, the
radius of the solid line shows the shape of the bubble, and is
proportional to its expansion speed.  The arrows indicate the
magnitude and direction of the flow in the shell.  The distance
between the dashed line and the solid line is proportional to the
surface density of the bubble.  The surface density jumps drastically
near the equator, indicating the presence of a dense disk.}
\end{figure}
\begin{figure}[tbp]
\epsfxsize=4.5in \centerline{\epsfbox{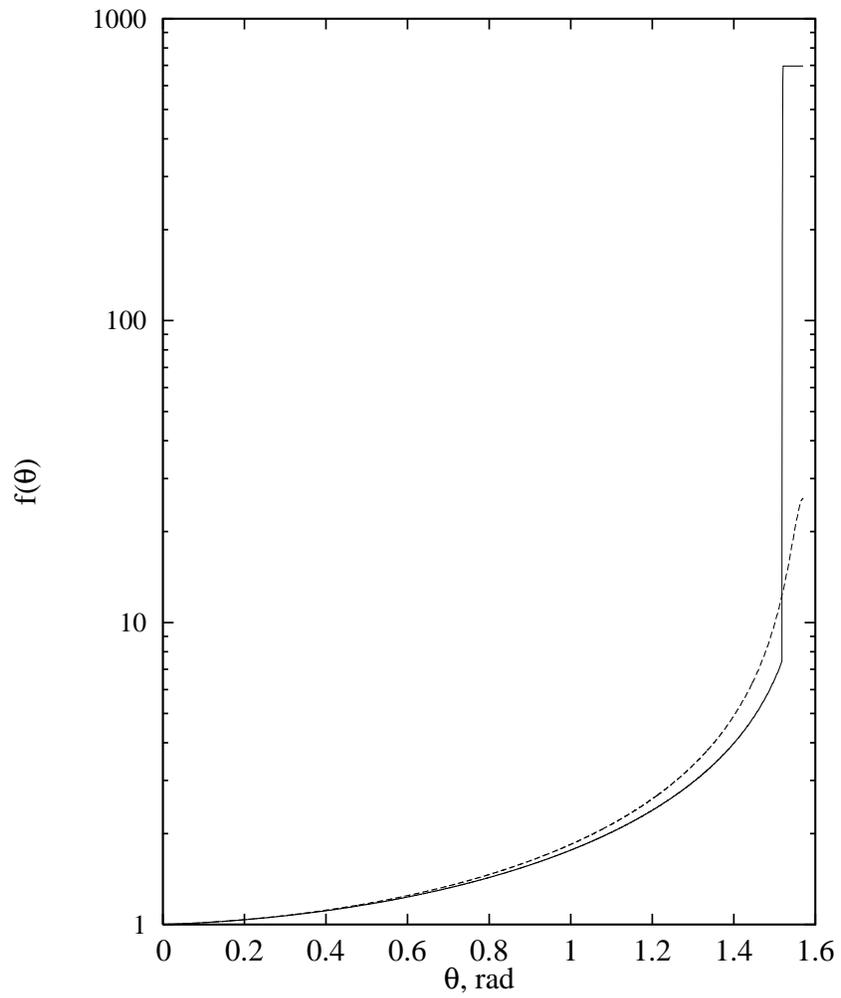}} \figcaption{The
density asymmetry functions for the RSG wind, for the models with
(solid) and without (dashed) a wind-compressed disk.}
\end{figure}
\end{document}